# Impact of Connected and Automated Vehicles on Transport Injustices

Laura Martínez-Buelvas*, Andry Rakotonirainy, Deanna Grant-Smith, and Oscar Oviedo-Trespalacios.

*Abstract*— Connected and automated vehicles (CAVs) are poised to transform the transport system. However, significant uncertainties remain about their impact, particularly regarding concerns that this advanced technology might exacerbate injustices, such as safety disparities for vulnerable road users (VRUs). Therefore, understanding the potential conflicts of this technology with societal values such as justice and safety is crucial for responsible implementation. To date, no research has focused on what safety and justice in transport mean in the context of CAV deployment and how the potential benefits of CAVs can be harnessed without exacerbating the existing vulnerabilities and injustices VRUs face. This paper addresses this gap by exploring car drivers' and pedestrians' perceptions of safety and justice issues that CAVs might exacerbate using an existing theoretical framework. Employing a qualitative approach, the study delves into the nuanced aspects of these concepts. Interviews were conducted with 30 participants (40% pedestrians) in Queensland, Australia, aged between 18 and 79. These interviews were recorded, transcribed, organised, and analysed using reflexive thematic analysis. Three main themes emerged from the participants' discussions: (1) CAVs as a safety problem for VRUs, (2) CAVs as a justice problem for VRUs, and (3) CAVs as an alignment with societal values problem. Participants emphasised the safety challenges CAVs pose for VRUs, highlighting the need for thorough evaluation and regulatory oversight. Concerns were also raised about CAVs potentially marginalising vulnerable groups within society. Participants advocated for inclusive discussions and a justice-oriented approach to designing a comprehensive transport system to address these concerns.

I. INTRODUCTION

Connected and Automated Vehicles (CAVs) are a transformative class of technology, capable of communicating with each other and with traffic infrastructure to navigate without human input. By reducing human performance variability, a primary cause of road accidents and fatalities, and alleviating traffic congestion, CAVs have the potential to significantly improve the efficiency and safety of land transport [1]. Notwithstanding this potential, one critical concern surrounding the deployment of CAVs is their potential to amplify persistent injustices within the transport system, especially regarding their interactions with vulnerable road users (VRUs). Pedestrians, lacking external protective devices to absorb the physical impact of a crash, face a heightened risk of becoming casualties in road accidents [2]. This vulnerability is reflected in global road trauma statistics, where pedestrians account for over 1.3 million annual fatalities, representing 26% of the total [3]. Consequently, pressing questions emerge: *what does safety and justice in transport mean in the context of CAV deployment*, and *how can the potential benefits of CAVs be harnessed without exacerbating the existing vulnerabilities and injustices faced by VRUs?*

Some progress has been made in understanding the unintended consequences of adopting CAVs, particularly concerning interactions with VRUs and managing social justice and sustainability impacts [4]. For example, [5] conducted a comparative analysis of the predictive capacities of the Theory of Planned Behaviour (TPB), the Technology Acceptance Model (TAM), and the Unified Theory of Acceptance and Use of Technology (UTAUT) regarding pedestrians' intentions to cross a road in front of a fully autonomous vehicle (AV). The results revealed that pedestrians' confidence in their ability to cross in front of an AV and their perception of approval from significant others played crucial roles in determining their intention to do so. Another study employed a statistical model known as a random parameters beta hurdle model to explore appropriate fines for pedestrians who intentionally block fully automated vehicles. Results indicated that age, gender, education level, violations, attitudes, behaviours promoting social interactions, and perceived ease or difficulty of interacting with fully automated vehicles would impact the likelihood and propensity of fines. The authors conclude that when formulating adoption strategies for fully automated vehicles in urban transport, transport agencies, federal governments, and stakeholders must address factors beyond legislation to ensure safe integration to prevent conflicts between pedestrians and fully automated vehicles [6]. Another research analysed equity implications in autonomous vehicle (AV) policies, categorising them into access, multimodal transportation, and community well-being. The authors advocated prioritising a shared AV model over the prevailing private ownership, emphasising positive sustainability

*Laura Martínez-Buelvas is with Queensland University of Technology (QUT), Centre for Accident Research and Road Safety – Queensland (CARRS-Q), QLD 4059, Australia; on leave from Universidad Tecnológica de Bolívar, Department of Industrial Engineering, Cartagena, Colombia (corresponding author e-mail: laurapatricia.martinezbuelvas@hdr.qut.edu.au – lmartinez@utb.edu.co)

Andry Rakotonirainy PhD is with Queensland University of Technology (QUT), Centre for Accident Research and Road Safety – Queensland (CARRS-Q), QLD 4059, Australia (e-mail: r.andry@qut.edu.au).

Deanna Grant-Smith PhD is with Queensland University of Technology (QUT), Faculty of Business & Law, School of Management, QLD 4000, Australia (e-mail: deanna.grantsmith@qut.edu.au)

Oscar Oviedo-Trespalacios PhD is with Delft University of Technology, Faculty of Technology, Policy and Management, Section of Safety and Security Science, 2628 BX Delft, The Netherlands (e-mail: O.OviedoTrespalacios@tudelft.nl)

implications such as incorporating electric energy sources to address racial disparities with AVs [7]. On the contrary, [8] examined the potential inequities CAVs may create for VRUs. The study developed a conceptual framework that considers current research and adoption trajectories of CAVs that can result in social injustices such as maintaining current road trauma risk for VRU, increase in pollution, traffic injuries, road responsibility, and space loss. Grasping these tensions is key to shedding light on issues regarding the impact of technology, which may cause its rejection or result in unforeseen outcomes or intended and perverse consequences.

CAVs are expected to be deployed in the coming years [9], and attention is now turning towards how CAVs should be regulated to facilitate a responsible adoption of the technology, considering alignment with broader societal issues such as safety, justice, and sustainability [10]. This research addresses a gap in justice-sensitive strategies by emphasising the importance of considering the potential impacts of CAVs and actively including the experiences of VRUs. For this reason, this paper evaluates the perceptions of pedestrians and car drivers about the safety and justice of CAVs deployment through an existing theoretical framework developed by [8]. We acknowledge that the impact of CAV deployment on justice in transport would remain uncertain if VRUs are not taken into account. Our research seeks to bridge this gap by engaging with road users in identifying critical concerns that policymakers and technology developers must address.

## II. METHOD

### A. Procedure

For this study, we employed a qualitative research method using semi-structured interviews with car drivers and pedestrians to explore their awareness of issues that the deployment of CAVs into level 5 of automation may create or exacerbate in the transport system. The decision to include pedestrians in the study is justified by their inherent vulnerability and frequent interactions with motorised vehicles. Additionally, incorporating car drivers into the research offers a comprehensive viewpoint, considering their paramount role in the transport system, the more logical market segment, and the potential influence on policy.

The conceptual framework *"A transport justice approach to integrating vulnerable road users with automated vehicles"* developed by [8] was used to guide the interview schedule. It focuses on three dimensions of transport justice: equality, fairness, and access. The procedure was the following: we examined pedestrians' and car drivers' perceptions of safety and justice within the existing transport infrastructure with questions such as, *what does justice/safety in transport mean for you* and *what needs to change in the transport system to improve justice/safety?* Additionally, we explored their awareness of any safety or justice issues that could arise or worsen with the introduction of CAVs. We presented examples of transport system inequities experienced by VRUs developed in the conceptual framework design by [8]. Then we asked questions such as, *what do you think about the transport justice problems CAVs will bring in the future* and, *now that you are aware of some of the problems associated with CAVs, is there anything else that worries you about them and their implementation?* Before delving into CAV-related topics, participants were queried about their familiarity with CAVs and whether they were aware of any potential future challenges posed by this technology. We provided a brief overview of what constitutes a CAV for those lacking knowledge.

Participants were interviewed face-to-face or via MS Teams for between 45 and 60 minutes. All interviews were audio-recorded and transcribed verbatim. Regarding the sample size, our desired data quality level drove our sample size selection following [11] proposing that semi-structured interviews are suitable for small sample sizes. Each participant received a $50 e-gift Australia card for participating in the study. The Research Ethics Committee approved this study (reference number 6593). Finally, data collection took place between February and April 2023.

### B. Participants

Pedestrians or car drivers aged 18 years or older residing in the greater Brisbane area (Queensland, Australia) who were willing to participate in a face-to-face or MS Teams interview were eligible for inclusion in this study. A total of 30 participants (18 car drivers and 12 pedestrians) agreed to participate. The car drivers had an age range of 20–79 years (M = 48.3, SD = 18.77), while the pedestrians were aged between 18 and 61 years (M = 36.0, SD = 12.94). On average, car drivers reported driving approximately 4.67 days per week, while pedestrians reported walking an average of 6.5 days per week. All participants' details are presented in Table I.

TABLE I. 'PARTICIPANTS' DETAILS

| Participant Code | Age (years) | Gender | Nationality |
|---|---|---|---|
| D1 | 61 | M | Australia |
| D2 | 79 | M | UK |
| D3 | 60 | F | Australia |
| D4 | 46 | M | Australia |
| D5 | 78 | F | Australia |
| D6 | 42 | M | China |
| D7 | 57 | F | Australia |
| D8 | 70 | F | Australia |
| D9 | 25 | NB | Australia |
| D10 | 37 | F | Australia |
| D11 | 26 | F | Canada |
| D12 | 53 | F | Australia |
| D13 | 20 | M | India |
| D14 | 58 | F | Australia |
| D15 | 34 | F | Australia |
| D16 | 36 | F | Australia |
| D17 | 20 | F | Australia |
| D18 | 68 | F | Australia |
| P1 | 61 | F | Australia |
| P2 | 28 | F | Australia |
| P3 | 29 | F | Iran |
| P4 | 23 | M | Philippines |
| P5 | 53 | M | Australia |
| P6 | 54 | F | Australia |
| P7 | 37 | F | Bangladesh |
| P8 | 18 | F | Nepal |
| P9 | 32 | F | Australia |
| P10 | 38 | F | Argentina |
| P11 | 24 | M | Australia |
| P12 | 35 | M | Australia |

## C. Data analysis

Reflective thematic analysis, a theoretically adaptable qualitative data interpretation method [12], was used to analyse the data collected from the interviews. The deductive, semantic, and critical realist/essential approach adopted facilitated pattern identification in the dataset [13]. Following [8]'s conceptual framework, we analysed the potential transport system inequities that VRUs could experience due to the deployment of CAVs in the future.

The co-authors engaged in discussions on coding, theme development, and interpretation of participant quotes, resolving differences through dialogue until a consensus was achieved. To maintain confidentiality, participant quotes are presented with pseudonyms, where "D" denotes driver and "P" denotes pedestrian.

## III. RESULTS

Three main themes synthesised participants' responses concerning safety and justice problems that may arise from wider CAV deployment. The themes were common among all participants, with no discernible variations based on age or gender. Identified differences were reported.

### A. CAVs as a safety problem for VRUs

CAVs can potentially reduce crashes and significantly enhance road safety by reducing human variability and bolstering safety measures. However, both participants concurred that safety remains a paramount concern in developing and implementing CAVs, as they present unique safety challenges for other road users, including VRUs such as pedestrians and cyclists. Consequently, the implementation of CAVs requires comprehensive evaluation and regulatory oversight. The consensus is that despite considerable technological advancements in automation, CAVs will be infallible. Indeed, participants expressed scepticism about the current state of CAV safety interactions with VRUs and infrastructure and the need for more research to provide confidence in their widespread adoption.

Drivers remarked,

*"There's no good evidence that there has been any improvement in safety or risk to people through injury."* [D1]

*"No automated system is foolproof, which is why I don't think we'll ever have fully automated cars that improve the safety of the VRUs."* [D5]

Relatedly, a pedestrian stated that ensuring the safety of VRUs should be a fundamental element of all CAV design and planning for their broader use:

*"I guess, before we can progress with them [CAVs], we should focus on just pedestrians and other kinds of vulnerable users' safety before they progress much further because there are yet significant issues."* [P2]

Many participants raised concerns about the real-world functionality of CAVs and distrust in the technology:

*"It makes me think that it's actually going to be even more challenging than I thought to make connected and automated vehicles work…there are a lot of assumptions being made about them, about the benefits that they might bring, and you know it's not."* [P12]

*"…it just doesn't seem real. There are too many things to consider when you're on the road. Is the automation going to be able to do all these moves that you have to make? Stop, or do pedestrians have to?"* [D18]

CAV interaction with the current infrastructure was also raised as a concern: *"I think the technology and infrastructure are years away from matching up to them being an effective prospect."* [D15]

Due to these concerns and uncertainties, additional research on CAV adoption is required to facilitate in-depth discussions and deliberations about their safety benefits and challenges in relation to addressing injustices regarding safety for VRUs:

*"I think that there'd have to be a lot more research and evidence. There has to be proof that they're safe. I'd have to be proof that they're safer than current ones."* [D5]

*"More research should be done because you should build that trust between the person and the technology before you fully engage with it. So, for as long as there's no data about this yet, then maybe I would have reservations using this."* [P4]

### B. CAVs as a justice problem for VRUs

Previous research has acknowledged the significant advantages of CAVs, such as enhanced accessibility to transport services for non-drivers or people with disabilities [14,15]. Despite this, many participants expressed concerns about the potential for CAVs to further exclude vulnerable population groups from participating in mainstream society, as many will not have access to or be able to afford high-end technology:

*"I would like to see these cars be available for people who need the most, which is those with disabilities, who are vision impaired and so on. But it makes me think that these cars would be available and affordable for the same people who can buy a Tesla car, you know? Rich people who already have lots of money and who want to use it."* [P6]

Many participants recognised the systemic injustices in the transport system and the need to ensure that all individuals, regardless of their background or identity, have equal access to the rights and benefits of CAV introduction.

*"It's really important that social justice is absolutely considered and accessibility for everyone. You know, coming back to what we talked about at the beginning, everyone should be able to get from A to B in a safe, pleasant, and accessible way."* [D7]

A greater understanding of the extent and characteristics of the impact of CAVs on the transport system for all road users remains needed. For example, one pedestrian expressed concern that reducing the space available for VRUs with the introduction of CAVs will cause injustices related to accessibility: *"I don't understand why they think that the congestion will be reduced by using these cars because these cars will take the same space as the regular way vehicles. So,*

*I don't find it too logical because they prefer a separate lane, like a separate spatial attention or special treatment. So, it will create more inequity."* [P7]

Finally, many participants expressed their desire to improve accessibility and quality of life for all by sharing CAVs or improving public transport access:

*"Free public transport and electrified public transport designed for everyone's accessibility would be the number one priority. I think they need vehicles with lots of people sharing them."* [D7]

*"I think the general aim should be how we can improve everyone's quality of life, especially for those who are the least benefited of all, like those who are disadvantaged."* [P10]

### C. CAVs as an alignment with societal values problem

Participants voiced apprehensions that the current trajectory of CAV development and deployment is predominantly steered by industry and private sector agendas. They fear these interests may need to align with the overarching objectives of sustainable development. This misalignment raises concerns about whether the benefits of CAVs are tailored more towards commercial gain than public welfare. These participants stress the importance of considering the broader societal implications of CAVs. They argue that integrating these technologies into our transport systems should focus on technical and economic feasibility and how they can enhance the social fabric. This includes ensuring equitable access to transport, reducing environmental impact, and improving the overall quality of urban life.

*"I also think that, of course, companies will advocate for this type of car, and they will always be saying good things, but because it is their business to sell you a car, we should be careful with that. Governments, on the one hand, want a society that thrives and that is modern, but sometimes they forget about the most basic things: people."* [P9]

*"I think that part of the policy of creating, you know, better cities is to ensure that many different groups and ages and people of diversity are involved in the process, So I think getting everybody else's opinion and experiences is important."* [P6]

The design and implementation of CAVs should prioritise reducing the carbon footprint of the transport, thus contributing to environmental sustainability. Additionally, deploying CAVs should be sensitive to the needs of all population segments, including those who may need direct access to these technologies. This could involve developing CAVs that enhance public transport systems or creating models that are affordable and accessible to a broader range of users.

*"Considering the environmental footprint price, I assume CAV would be electric. I mean, if I get some statistics or some data, then I could make a better option of judgment to buy one."* [D13]

*"I would like to be able to, you know, if these cars would also help to reduce emissions."* [P6]

### D. A thematic roadmap

The analysis identified three key themes based on participants' awareness of the justice problems and safety challenges created or potentially exacerbated by the deployment of CAVs. These themes include CAVs as a safety issue for VRUs, CAVs as a justice problem for VRUs, and CAVs as an alignment with societal values problem. Concerns were voiced regarding CAVs potentially excluding vulnerable groups from society. To ensure the successful integration of CAVs into our transport system, participants advocated for centred on inclusive discussions and a justice-oriented approach to comprehensive transport system design.

Fig 1. presents a thematic framework to visually represent these themes and their interrelationships with the conceptual transport justice framework presented by [8]. Fig 1 maps current or potential injustices experienced by VRUs that CAVs would exacerbate in the future against the three dimensions of transport justice: equality, fairness, and access, following the conceptual framework developed by [8]. According to this framework and the findings of this study, CAVs pose safety risks to VRUs due to their preliminary technology development, increased crashes, shifting road responsibilities, and the necessity for accessible warning systems to protect VRUs. Also, CAVs would create justice issues for VRUs by reducing on-street space and potentially limiting access to services and inclusion for disabled and elderly individuals. Finally, CAVs pose societal value challenges by affecting access, autonomy, and inclusion for disabled and elderly individuals, exacerbating congestion and pollution, and altering road responsibilities between CAVs and VRUs.

## IV. DISCUSSION

The present research found that CAVs are marked by significant scepticism regarding their purported safety benefits. This scepticism largely stems from concerns about their safety and real-world functionality. Arguably, this can be explained by unethical industry practices that have misled the public concerning the automation capabilities of vehicles in the market [16] and major safety-critical incidents with VRUs [17]. Furthermore, there's a call for more rigorous research and evidence to validate the safety claims of CAVs.

Participants underscored the need for careful consideration of social justice and equitable distribution of technology benefits in implementing CAVs. These concerns somewhat align with previous research highlighting the potential benefits of CAVs, such as improved access to transport services for non-drivers or people with disabilities [18-20]. Participants also highlighted the importance of social justice in CAV deployment, stressing the need for equitable distribution of resources and opportunities. There's a recognition that the introduction of CAVs should not only focus on technological advancement but also ensure that all individuals, regardless of their socioeconomic background, have equal access to the benefits brought by these vehicles.

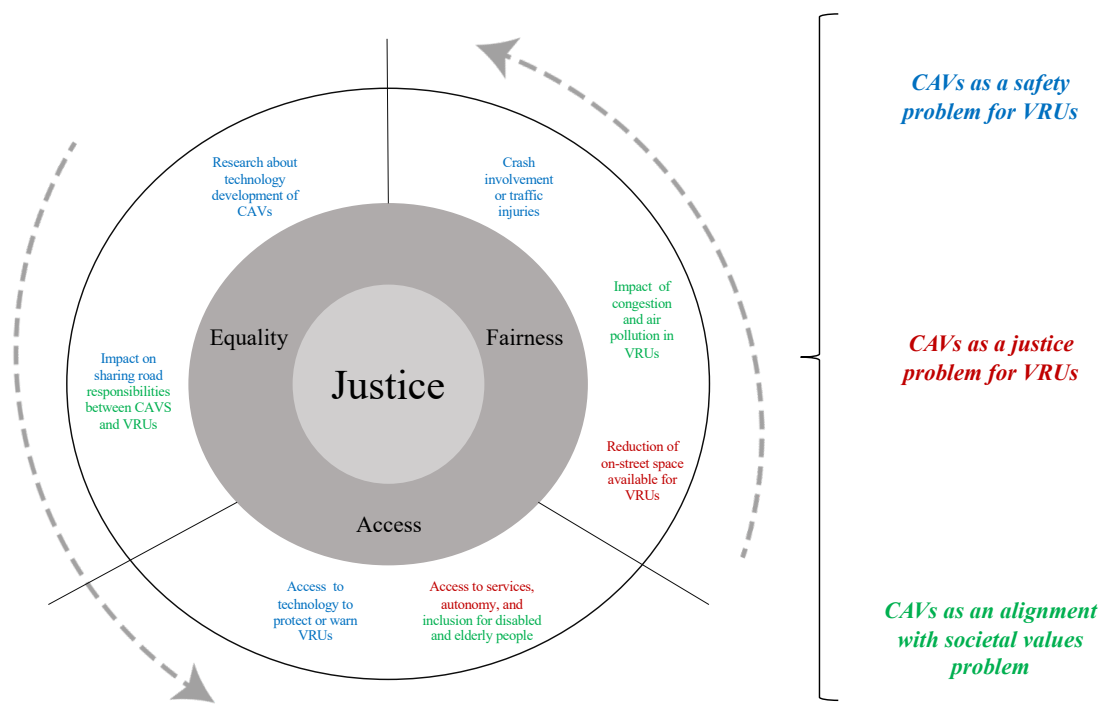

Figure 1. A thematic roadmap contrasting the model presented by [8] and our findings

This indicates a need for policies prioritising community well-being and equitable access rather than focusing solely on supporting the introduction of a single technology like CAVs.

Participants underscored the necessity of involving the community in discussions about the design and implementation of CAVs to prevent potential misuse by the industry. They advocate for a proactive and future-centric approach in developing and deploying CAVs, which involves aligning these vehicles with anticipated long-term societal changes. For instance, this could include designing CAVs to integrate with advanced smart city infrastructures, such as traffic systems that adapt to varying road conditions in real time. An explicit example would be if CAVs were equipped with sensors and communication systems that interact with intelligent traffic lights to optimise traffic flow and reduce congestion.

In addition, preparing for future environmental regulations is crucial, like ensuring CAVs are designed for easy upgrades to meet stricter emissions standards or to incorporate renewable energy technologies. Anticipating urbanisation trends, CAVs could be designed with space-saving features for densely populated areas, such as foldable structures for compact parking or advanced safety systems to protect pedestrians in crowded urban settings. This forward-thinking approach ensures that CAVs are technologically advanced and remain relevant, beneficial, and sustainable within a dynamically changing societal context.

## V. CONCLUSION

This research emphasises the urgent requirement for policymakers and practitioners within the Connected and Automated Vehicles (CAVs) industry to tackle the prevalent doubts about CAVs' safety advantages, especially concerning protecting vulnerable road users. The resolution to this doubt is to guarantee that the design and deployment of CAVs are both transparent and ethical, bolstered by thorough scientific research and evidence. Furthermore, it is critical to pivot towards policies that propel technological progress, champion social justice, and ensure equitable accessibility across all population segments, thereby integrating technological progress with broader social objectives, including sustainability.

Embracing a comprehensive strategy that melds safety research, ethical standards, and social equity with thoughtful strategic planning is pivotal. This strategy aims to guarantee that the benefits derived from CAVs are shared fairly and have a beneficial impact on the well-being of communities. This study propels the conversation about CAVs forward by integrating technological innovations with societal necessities, presenting a broad view that covers safety, ethical considerations, and the enduring welfare of society. These insights prove to be essential for policymakers, practitioners, and developers advocating for a cohesive strategy that synchronises CAVs with societal aspirations.

Lastly, a well-recognised limitation of this qualitative investigation, which is common to studies of this nature, is its emphasis on depth over breadth. This methodological choice inherently restricts the capacity to generalise the findings to broader populations. Furthermore, the potential for researcher subjectivity to influence both the analysis and interpretation of data introduces an additional layer of complexity, underscoring the necessity of acknowledging these factors when evaluating the study's contributions and implications. This qualitative analysis raises significant questions about developing and implementing technology, particularly making CAVs more attuned to societal needs. Readers with a technical background or those with a keen interest in the technological aspects will find that the insights offered challenge existing paradigms and encourage re-evaluating how CAV technologies can be designed and deployed to serve better and reflect societal priorities and values. This perspective is crucial for ensuring that technological advancements are socially responsible and aligned with the broader equity and sustainability goals.

ACKNOWLEDGMENT

This research is funded by iMOVE CRC and supported by the Cooperative Research Centres program, an Australian Government initiative [Project code: 5-006]. Laura Martínez-Buelvas and Andry Rakotonirainy received support from the Motor Accident Insurance Commission (MAIC) Queensland. The funders had no role in the study design, data collection and analysis, publication decision, or manuscript preparation.